\documentclass[prd,aps,preprint,amsmath,nofootinbib,amssymb,eqsecnum,showkeys]{revtex4-1}
\pdfoutput=1
\usepackage{subfigure}
\usepackage{graphicx}
\usepackage{xcolor}
\newcommand{\be}{\begin{equation}}
\newcommand{\ee}{\end{equation}}
\newcommand{\bea}{\begin{eqnarray}}
\newcommand{\eea}{\end{eqnarray}}

\begin{document}
\title{Exploring the dark annihilation: multi-component asymmetric and symmetric dark matter}

\author{Amit Dutta Banik}
\email[E-mail: ]{amitdbanik@gmail.com}
\affiliation{Physics and Applied Mathematics Unit, Indian Statistical Institute, Kolkata-700108, India}

\date{\today}

\begin{abstract}
The article describes Boltzmann equations for a potential case of multi-particle dark matter with symmetric and asymmetric dark matter components in a model-independent approach. We focus on the specific scenario where one of the DM candidates remains completely invisible, having only hidden sector interactions with the other dark matter constituent referred to as ``dark annihilation". 
The possible effect of non-standard expansion of the universe on the dark matter abundance is also taken into account.
  
\end{abstract}


\maketitle

\section {Introduction}

Experimental findings have firmly established that more than 80\% of total matter density is shared by unknown dark matter (DM) that challenges the Standard Model (SM) of particle physics for decades, broadening the path beyond SM phenomenology~\cite{Aghanim:2018eyx}. Among the many proposed candidates based on their thermal behaviour, and origin, weakly interacting massive particles (WIMPs) remain the most favoured DM candidate investigated via various direct and indirect search experiments~\cite{Jungman:1995df,Bertone:2004pz}. Annihilation of WIMPs at the electroweak scale naturally provides the correct DM relic abundance, known as the WIMP miracle. However, WIMPs, to date, remain elusive towards detection via direct and indirect detection experiments and invoke the avenue for alternate theories of DM origin and its property.    

The standard annihilation of WIMPs demands the presence of a DM particle and its antiparticle of equal density. However, as we observe asymmetry in the visible matter density, the well-known baryon asymmetry in the Universe, it is plausible that there exists an asymmetry between WIMP particle and anti-particle density. Such scenarios have been predicted in theories as alternative of WIMPs as asymmetric dark matter (ADM)~\cite{Kaplan:2009ag,Graesser:2011wi,Iminniyaz:2011yp,Gelmini:2013awa,Zurek:2013wia,Kitabayashi:2015oda,Blennow:2015xha,Iminniyaz:2016iom,Agrawal:2016uwf,Nagata:2016knk,Baldes:2017gzw,Gresham:2017cvl,HajiSadeghi:2017zrl}, have recently gained interest in phenomenological and observational aspects~\cite{Murase:2016nwx,Gresham:2018anj,Lopes:2019jca,Ghosh:2020lma,Ghosh:2021qbo,Ho:2022erb,Steigerwald:2022pjo,Ho:2022tbw,Roy:2024ear}. Although, many phenomenological models of WIMP have been discussed in literature, whether those are symmetric or asymmetric in nature, or some other variants, the concept of multiple WIMP have been overlooked and has received attention recently in past decade ~\cite{Belanger:2011ww,Bhattacharya:2013hva,Bian:2013wna,Esch:2014jpa,Ahmed:2017dbb,Herrero-Garcia:2017vrl,Herrero-Garcia:2018qnz,Aoki:2018gjf,Bhattacharya:2016ysw,Chakraborti:2018aae,Elahi:2019jeo,Borah:2019aeq,Bhattacharya:2019tqq,DuttaBanik:2020jrj,BasiBeneito:2022qxd,DuttaBanik:2023yxj}. In this work, we consider a case of multi-component dark matter with an asymmetric DM component and a symmetric DM component. Both the DM components are stabilised by some symmetry and depending on the situation, one candidate can annihilate into the other, leading to novel coupled Boltzmann equations (BEs) between asymmetric and symmetric DM\footnote{
It is to be noted that the symmetric DM is not the symmetric component of the asymmetric dark matter, considered in conventional ADM~\cite{Graesser:2011wi}.}. One interesting situation occurs when, in particular, one candidate annihilates into the hidden sector only, being invisible to the visible sector or the ``dark annihilation'' dominates over SM sector annihilation of the DM candidate. This enlightens a completely new phenomenology of the multi-particle dark matter sector.

Solving coupled Boltzmann equations, relic densities of DM candidates are obtained in the standard radiation-dominated era.
However, the effects of non-standard cosmological history can change the abundance of DM components significantly. Therefore, it is also interesting to consider such effects on the abundances of both asymmetric and symmetric DM candidates. We consider a very well established model for non-standard expansion of universe with a scalar field and study the effects of modified cosmology in the present context ~\cite{DEramo:2017gpl,DEramo:2017ecx,Biswas:2018iny,Fernandez:2018tfa,Chen:2019etb,Mahanta:2019sfo,Konar:2020vuu,Barman:2018esi,Chakraborty:2022gob}. 

The paper is organised as follows. We begin with the formulation of the Boltzmann equation for the multi-particle dark matter with an asymmetric DM candidate accompanied by a symmetric component of DM in Sec. II. The nonstandard thermal history of the universe and the modified BEs are discussed later in Sec. III. Calculation of DM relic abundance is presented in Sec. IV and finally the paper is concluded in Sec. V.

\section{Boltzmann equations with coupled asymmetric and symmetric WIMP dark matter}
\label{BE}
In this section, we formulate the Boltzmann equations (BEs) for multi-component WIMPs with a symmetric DM candidate and an asymmetric component. We refrain from the discussion of the origin of asymmetry and consider a simple two-component DM scenario with a complex scalar DM candidate $A$ with its complex conjugate $\bar{A}$ and a symmetric DM candidate $S$. Depending on the masses of DM particles we derive BEs for the case A) $m_A>m_S$ and B) $m_A<m_S$. We primarily keep the analysis to be model-independent and assume both the dark matter candidates are stable realised by some unbroken symmetry. 
\subsection{$m_{A}>m_{S}$} 
We write down the standard Boltzmann equation with asymmetric WIMP candidate ($A,\bar{A}$) and symmetric WIMP $S$ having annihilation into SM sector ($A\bar{A}\rightarrow XX$, $SS\rightarrow XX$) and annihilation into dark sector $A\bar{A}\rightarrow SS$ for $m_{A,\bar{A}}>m_S$.

\begin{align} \label{eq:boltzmann}
  \frac{Hx}{s}\frac{dY_A}{dx}&= -{\langle \sigma v \rangle}_{A\bar{A} \rightarrow XX} \left(Y_AY_{\bar{A}}-Y_{A,\rm eq}Y_{\bar{A},\rm eq}\right) - {\langle \sigma v \rangle}_{A\bar{A} \rightarrow SS} \left(Y_AY_{\bar{A}}-\frac{Y_{A,\rm eq}Y_{\bar{A},\rm eq}}{Y_{S,\rm eq}^{2}}Y_S^2\right)
\nonumber\\
  \frac{Hx}{s}\frac{dY_{\bar{A}}}{dx}&= -{\langle \sigma v \rangle}_{A\bar{A} \rightarrow XX} \left(Y_AY_{\bar{A}}-Y_{A,\rm eq}Y_{\bar{A},\rm eq}\right) - {\langle \sigma v \rangle}_{A\bar{A} \rightarrow SS} \left(Y_AY_{\bar{A}}-\frac{Y_{A,\rm eq}Y_{\bar{A},\rm eq}}{Y_{S,\rm eq}^{2}}Y_S^2\right)
\nonumber\\
 \frac{Hx}{s}\frac{dY_S}{dx} &= -{\langle \sigma v \rangle}_{S S \rightarrow XX} \left(Y_S^2-Y_{S,\rm eq}^2\right)
+ 2{\langle \sigma v \rangle}_{A\bar{A} \rightarrow SS} \left(Y_AY_{\bar{A}}-\frac{Y_{A,\rm eq}Y_{\bar{A},\rm eq}}{Y_{S,\rm eq}^{2}}Y_S^2\right),
\end{align}
where $Y_i=n_i/s$ with $n_i$ being the comoving number density and $s$ being the entropy, $x=m_A/T$ and $H$ is the Hubble rate. $Y_{i, \rm eq}=n_{i,\rm eq}/s$ denotes comoving number density in equilibrium, with $n_{i, \rm eq}$ being equilibrium number density, given as

\begin{align}\label{neq}
n_{A,\rm eq}=g_{A}\left(\frac{m_{A}T}{2\pi}\right)^{3/2}e^{(-m_A+\mu_{A})/T},\nonumber \\
n_{\bar{A},\rm eq}=g_{A}\left(\frac{m_{A}T}{2\pi}\right)^{3/2}e^{(-m_A-\mu_{A})/T},\nonumber \\
n_{S,\rm eq}=g_{S}\left(\frac{m_{S}T}{2\pi}\right)^{3/2}e^{-m_S/T}\, .
\end{align}

where $g_i$ are the internal degrees of freedom of particles and the mass of the particles is $m_i$.
The chemical potential of particle and anti-particle of asymmetric dark matter satisfies the relation $\mu_A=-\mu_{\bar{A}}$, and for the symmetric dark matter $\mu_S=0$. Looking into the first two equations of Eq.~(\ref{eq:boltzmann}) for asymmetric dark matter, it is easily observed that $\frac{d(Y_{A}-Y_{\bar{A}})}{dx}=0$, and $Y_{A}-Y_{\bar{A}}=C$ is constant.
With that, one can redefine the Boltzmann equations as

\begin{align} \label{eq:boltzmann1}
  \frac{Hx}{s}\frac{dY_A}{dx}&= -{\langle \sigma v \rangle}_{A\bar{A} \rightarrow XX} \left(Y_A^2-CY_A-Y_{A,\rm eq}Y_{\bar{A},\rm eq}\right) - {\langle \sigma v \rangle}_{A\bar{A} \rightarrow SS} \left(Y_A^2- CY_{A}-\frac{Y_{A,\rm eq}Y_{\bar{A},\rm eq}}{Y_{S,\rm eq}^{2}}Y_S^2\right)\, ,
\nonumber\\
  \frac{Hx}{s}\frac{dY_{\bar{A}}}{dx}&= -{\langle \sigma v \rangle}_{A\bar{A} \rightarrow XX} \left(Y_{\bar{A}}^2+CY_{\bar{A}}-Y_{A,\rm eq}Y_{\bar{A},\rm eq}\right) - {\langle \sigma v \rangle}_{A\bar{A} \rightarrow SS} \left(Y_{\bar{A}}^2+CY_{\bar{A}}-\frac{Y_{A,\rm eq}Y_{\bar{A},\rm eq}}{Y_{S,\rm eq}^{2}}Y_S^2\right)\, ,
\nonumber\\
 \frac{Hx}{s}\frac{dY_S}{dx} &= -{\langle \sigma v \rangle}_{S S \rightarrow XX} \left(Y_S^2-Y_{S,\rm eq}^2\right)
+ {\langle \sigma v \rangle}_{A\bar{A} \rightarrow SS} \left(Y_A^2- CY_{A}-\frac{Y_{A,\rm eq}Y_{\bar{A},\rm eq}}{Y_{S,\rm eq}^{2}}Y_S^2\right)
\nonumber\\
& \qquad +{\langle \sigma v \rangle}_{A\bar{A} \rightarrow SS} \left(Y_{\bar{A}}^2+CY_{\bar{A}}-\frac{Y_{A,\rm eq}Y_{\bar{A},\rm eq}}{Y_{S,\rm eq}^{2}}Y_S^2\right). 
\end{align}

The standard evolution of asymmetric dark matter with DM candidates annihilating into SM has been extensively studied in many literature.
As we focus our attention on multi-component dynamics with $A,\bar{A}$ and $S$, we find it convenient to drop the $A\bar{A}\rightarrow XX$ term considering the dark sector annihilation $A\bar{A}\rightarrow SS$ to be the dominant mode of asymmetric dark matter annihilation. This indicates that the freeze-out of the asymmetric DM is governed by dark or hidden sector annihilation only.
Therefore, we end up with the simplified BEs

\begin{align} \label{eq:boltzmann2}
 \frac{Hx}{s}\frac{dY_A}{dx}&= - {\langle \sigma v \rangle}_{A\bar{A} \rightarrow SS} \left(Y_A^2- CY_{A}-\frac{Y_{A,\rm eq}Y_{\bar{A},\rm eq}}{Y_{S,\rm eq}^{2}}Y_S^2\right)\, ,
\nonumber\\
\frac{Hx}{s}\frac{dY_{\bar{A}}}{dx}&= - {\langle \sigma v \rangle}_{A\bar{A} \rightarrow SS} \left(Y_{\bar{A}}^2+CY_{\bar{A}}-\frac{Y_{A,\rm eq}Y_{\bar{A},\rm eq}}{Y_{S,\rm eq}^{2}}Y_S^2\right)\, ,
\nonumber\\
 \frac{Hx}{s}\frac{dY_S}{dx} &= -{\langle \sigma v \rangle}_{S S \rightarrow XX} \left(Y_S^2-Y_{S,\rm eq}^2\right)
+ {\langle \sigma v \rangle}_{A\bar{A} \rightarrow SS} \left(Y_A^2- CY_{A}-\frac{Y_{A,\rm eq}Y_{\bar{A},\rm eq}}{Y_{S,\rm eq}^{2}}Y_S^2\right)
\nonumber\\
& \qquad +{\langle \sigma v \rangle}_{A\bar{A} \rightarrow SS} \left(Y_{\bar{A}}^2+CY_{\bar{A}}-\frac{Y_{A,\rm eq}Y_{\bar{A},\rm eq}}{Y_{S,\rm eq}^{2}}Y_S^2\right). 
\end{align}

The Eq.~(\ref{eq:boltzmann2}), satisfies the condition $\frac{d(Y_{A}-Y_{\bar{A}})}{dx}=0$. The chemical potential $\mu_{A}$ can be realised in terms of the asymmetry parameter $C$ of the Boltzmann equation \cite{Gelmini:2013awa},
\begin{equation}
e^{\mu_A /T}=\frac{1}{2}\left(\frac{Cs}{n_{A,\rm eq}(\mu =0)}+\sqrt{4+\left(\frac{Cs}{n_{A,\rm eq}(\mu =0)}\right)^2}\right).
\label{muandC}
\end{equation}

\subsection{$m_{A}<m_{S}$} 

We now delve into the scenario where the symmetric DM candidate is heavier allowing its annihilation into asymmetric DM candidates. This describes a scenario complementary to the previous case as the symmetric DM candidate $S$ annihilates into the asymmetric DM and undergoes annihilation
$SS\rightarrow A\bar{A}$.
Therefore, the Boltzmann equations are as follows 

\begin{align} \label{eq:boltzmann3}
  \frac{Hx}{s}\frac{dY_A}{dx}&= -{\langle \sigma v \rangle}_{A\bar{A} \rightarrow XX} \left(Y_AY_{\bar{A}}-Y_{A,\rm eq}Y_{\bar{A},\rm eq}\right) + {\langle \sigma v \rangle}_{SS \rightarrow A\bar{A}} \left(Y_S^2-\frac{Y_AY_{\bar{A}}}{Y_{A,\rm eq}Y_{\bar{A},\rm eq}}Y_{S,\rm eq}^{2}\right)\, ,
\nonumber\\
  \frac{Hx}{s}\frac{dY_{\bar{A}}}{dx}&= -{\langle \sigma v \rangle}_{A\bar{A} \rightarrow XX} \left(Y_AY_{\bar{A}}-Y_{A,\rm eq}Y_{\bar{A},\rm eq}\right) + {\langle \sigma v \rangle}_{SS \rightarrow A\bar{A}} \left(Y_S^2-\frac{Y_AY_{\bar{A}}}{Y_{A,\rm eq}Y_{\bar{A},\rm eq}}Y_{S,\rm eq}^{2}\right)\, ,
\nonumber\\
 \frac{Hx}{s}\frac{dY_S}{dx} &= -{\langle \sigma v \rangle}_{S S \rightarrow XX} \left(Y_S^2-Y_{S,\rm eq}^2\right)
-{\langle \sigma v \rangle}_{SS \rightarrow A\bar{A}} \left(Y_S^2-\frac{Y_AY_{\bar{A}}}{Y_{A,\rm eq}Y_{\bar{A},\rm eq}}Y_{S,\rm eq}^{2}\right)
\nonumber\\
& \qquad -{\langle \sigma v \rangle}_{SS \rightarrow A\bar{A}} \left(Y_S^2-\frac{Y_AY_{\bar{A}}}{Y_{A,\rm eq}Y_{\bar{A},\rm eq}}Y_{S,\rm eq}^{2}\right), 
\end{align}

where $x=m_S/T$. Eq.~(\ref{eq:boltzmann3}) also respects the condition $\frac{d(Y_{A}-Y_{\bar{A}})}{dx}=0$.
Therefore, using the relation $Y_{A}-Y_{\bar{A}}=C$, we obtain

\begin{align} \label{eq:boltzmann4}
  \frac{Hx}{s}\frac{dY_A}{dx}&= -{\langle \sigma v \rangle}_{A\bar{A} \rightarrow XX} \left(Y_A^2-CY_A-Y_{A,\rm eq}Y_{\bar{A},\rm eq}\right) + {\langle \sigma v \rangle}_{SS \rightarrow A\bar{A}} \left(Y_S^2-\frac{Y_A^2-CY_A}{Y_{A,\rm eq}Y_{\bar{A},\rm eq}}Y_{S,\rm eq}^{2}\right)\, ,
\nonumber\\
  \frac{Hx}{s}\frac{dY_{\bar{A}}}{dx}&= -{\langle \sigma v \rangle}_{A\bar{A} \rightarrow XX} \left(Y_{\bar{A}}^2+CY_{\bar{A}}-Y_{A,\rm eq}Y_{\bar{A},\rm eq}\right) + {\langle \sigma v \rangle}_{SS \rightarrow A\bar{A}} \left(Y_S^2-\frac{Y_{\bar{A}}^2+CY_{\bar{A}}}{Y_{A,\rm eq}Y_{\bar{A},\rm eq}}Y_{S,\rm eq}^{2}\right)\, ,
\nonumber\\
 \frac{Hx}{s}\frac{dY_S}{dx} &= -{\langle \sigma v \rangle}_{S S \rightarrow XX} \left(Y_S^2-Y_{S,\rm eq}^2\right)
-{\langle \sigma v \rangle}_{SS \rightarrow A\bar{A}} \left(Y_S^2-\frac{Y_{\bar{A}}^2+CY_{\bar{A}}}{Y_{A,\rm eq}Y_{\bar{A},\rm eq}}Y_{S,\rm eq}^{2}\right)
\nonumber\\
& \qquad -{\langle \sigma v \rangle}_{SS \rightarrow A\bar{A}} \left(Y_S^2-\frac{Y_A^2-CY_A}{Y_{A,\rm eq}Y_{\bar{A},\rm eq}}Y_{S,\rm eq}^{2}\right). 
\end{align}

Again, as we consider the hidden sector perspectives, motivated by the scenario of dark sector dynamics, we simply drop the standard model annihilation term of the symmetric dark matter $S$, assuming it only interacts with its own kind, i.e.; $A$ and $\bar{A}$. Therefore, the DM candidate $S$ avoids all interactions with the SM sector. This kind of scenario can also be achieved in models where dark sector coupling to SM is forbidden via some symmetry or cases where DM coupling to SM falls in the blind spot. However, as we do not intend to discuss any specific model, we solve for the BEs to obtain relic abundances of DM candidates in a model-independent way. Therefore, the BEs are now written as

\begin{align} \label{eq:boltzmann5}
  \frac{Hx}{s}\frac{dY_A}{dx}&= -{\langle \sigma v \rangle}_{A\bar{A} \rightarrow XX} \left(Y_A^2-CY_A-Y_{A,\rm eq}Y_{\bar{A},\rm eq}\right) + {\langle \sigma v \rangle}_{SS \rightarrow A\bar{A}} \left(Y_S^2-\frac{Y_A^2-CY_A}{Y_{A,\rm eq}Y_{\bar{A},\rm eq}}Y_{S,\rm eq}^{2}\right)\, ,
\nonumber\\
  \frac{Hx}{s}\frac{dY_{\bar{A}}}{dx}&= -{\langle \sigma v \rangle}_{A\bar{A} \rightarrow XX} \left(Y_{\bar{A}}^2+CY_{\bar{A}}-Y_{A,\rm eq}Y_{\bar{A},\rm eq}\right) + {\langle \sigma v \rangle}_{SS \rightarrow A\bar{A}} \left(Y_S^2-\frac{Y_{\bar{A}}^2+CY_{\bar{A}}}{Y_{A,\rm eq}Y_{\bar{A},\rm eq}}Y_{S,\rm eq}^{2}\right)\, ,
\nonumber\\
 \frac{Hx}{s}\frac{dY_S}{dx} &= 
-{\langle \sigma v \rangle}_{SS \rightarrow A\bar{A}} \left(Y_S^2-\frac{Y_{\bar{A}}^2+CY_{\bar{A}}}{Y_{A,\rm eq}Y_{\bar{A},\rm eq}}Y_{S,\rm eq}^{2}\right)
-{\langle \sigma v \rangle}_{SS \rightarrow A\bar{A}} \left(Y_S^2-\frac{Y_A^2-CY_A}{Y_{A,\rm eq}Y_{\bar{A},\rm eq}}Y_{S,\rm eq}^{2}\right). 
\end{align}

With the above specifications of dark sector in Eq.~(\ref{eq:boltzmann2},~\ref{eq:boltzmann5}), the multi-component dark matter scenario described, results in a dark matter candidate that interacts (annihilates) into visible sector and decouples from the thermal bath after freeze-out while the other candidate remains hidden, decouples from the thermal bath as the dark annihilation freezes out.

\section{Non-Standard Cosmological era}
\label{NS}
In Sec.~\ref{BE}, we have presented the Boltzmann equations for a two-component asymmetric and symmetric dark matter focusing on the hidden sector annihilation of the heavier DM while the lighter DM annihilates into SM sector. BEs derived in Eq.~(\ref{eq:boltzmann2}) and Eq.~(\ref{eq:boltzmann5}) are valid when the universe is radiation-dominated. However, the universe may not be radiation-dominated at the early stage. For example, one can consider a scalar field $\phi$ having energy density $\rho_{\phi}$ that varies with scale factor as $\rho_{\phi}\simeq a^{-(4+n)}$; $n>0$ \cite{DEramo:2017gpl}. Therefore at the early age of post inflationary universe after reheating, the energy density of $\phi$ field dominates over standard radiation domination (RD)
as $\rho_{\rm RD}\simeq a^{-4}$. As a consequence, the universe expands faster than expected when the scalar $\phi$ remains active, and the Hubble parameter is modified $H\propto \sqrt{\rho_{\rm RD}+\rho_\phi}$, where $\rho_{\rm RD}=\frac{\pi^2}{30}g_*T^4$, and $g_*$ is the effective relativistic degrees of freedom.
With the assumption that the field $\phi$ couples to gravity only, and entropy of the universe remains constant $S=sa^3$, we define a reference temperature $T_R$ such that at $T=T_R$, $\rho_{\phi}=\rho_{\rm RD}$.
Therefore, energy density of $\phi$ can be expressed as \cite{DEramo:2017gpl}
\be
\rho_{\phi}(T)=\rho_{\rm{RD}}(T_R)\bigg[\bigg(\frac{g_{*s}(T)}{g_{*s}(T_R)}\bigg)^{(4+n)/3}\bigg(\frac{T}{T_R}\bigg)^{4+n}\bigg]\,\, .
\label{rho-phi}
\ee
At higher temperatures, all the particle species are in equilibrium, and for $T_R\simeq$ GeV $g_{*s}(T_R)\equiv g_{*s}(T)$, and $g_*(T)\equiv g_*(T_R)$. This leads to the modified Hubble parameter at temperature $T$ as
\be
H^{\prime}=H\left[1+\left(\frac{T}{T_R}\right)^n\right]^{1/2},
\ee
where $H=\sqrt{\frac{g_*}{90}}\frac{\pi T^2}{M_P}$, $M_P=2.4\times 10^{18}$ GeV, is the reduced Planck mass. Therefore, if the scalar field is active, the expansion of the universe will be defined by $H^{\prime}$, and it will affect the evolution of multi-component dark matter Boltzmann equations derived in Sec.~\ref{BE}.

\subsection{BEs in NS Cosmology}
As discussed above, the non-standard expansion of the universe in the early universe can affect the dark matter abundance
if the scalar fields remain effective before the freeze-out of dark matter particles. As a result, Boltzmann equations for multi-particle DM system described in Sec.~\ref{BE} are redefined as

\begin{align} \label{eq:boltzmann6}
 \frac{H^{\prime}x}{s}\frac{dY_A}{dx}&= - {\langle \sigma v \rangle}_{A\bar{A} \rightarrow SS} \left(Y_A^2- CY_{A}-\frac{Y_{A,\rm eq}Y_{\bar{A},\rm eq}}{Y_{S,\rm eq}^{2}}Y_S^2\right)\, ,
\nonumber\\
\frac{H^{\prime}x}{s}\frac{dY_{\bar{A}}}{dx}&= - {\langle \sigma v \rangle}_{A\bar{A} \rightarrow SS} \left(Y_{\bar{A}}^2+CY_{\bar{A}}-\frac{Y_{A,\rm eq}Y_{\bar{A},\rm eq}}{Y_{S,\rm eq}^{2}}Y_S^2\right)\, ,
\nonumber\\
 \frac{H^{\prime}x}{s}\frac{dY_S}{dx} &= -{\langle \sigma v \rangle}_{S S \rightarrow XX} \left(Y_S^2-Y_{S,\rm eq}^2\right)
+ {\langle \sigma v \rangle}_{A\bar{A} \rightarrow SS} \left(Y_A^2- CY_{A}-\frac{Y_{A,\rm eq}Y_{\bar{A},\rm eq}}{Y_{S,\rm eq}^{2}}Y_S^2\right)
\nonumber\\
& \qquad +{\langle \sigma v \rangle}_{A\bar{A} \rightarrow SS} \left(Y_{\bar{A}}^2+CY_{\bar{A}}-\frac{Y_{A,\rm eq}Y_{\bar{A},\rm eq}}{Y_{S,\rm eq}^{2}}Y_S^2\right). 
\end{align}

for $m_{A}>m_{S}$, and 

\begin{align} \label{eq:boltzmann7}
  \frac{H^{\prime}x}{s}\frac{dY_A}{dx}&= -{\langle \sigma v \rangle}_{A\bar{A} \rightarrow XX} \left(Y_A^2-CY_A-Y_{A,\rm eq}Y_{\bar{A},\rm eq}\right) + {\langle \sigma v \rangle}_{SS \rightarrow A\bar{A}} \left(Y_S^2-\frac{Y_A^2-CY_A}{Y_{A,\rm eq}Y_{\bar{A},\rm eq}}Y_{S,\rm eq}^{2}\right)\, ,
\nonumber\\
  \frac{H^{\prime}x}{s}\frac{dY_{\bar{A}}}{dx}&= -{\langle \sigma v \rangle}_{A\bar{A} \rightarrow XX} \left(Y_{\bar{A}}^2+CY_{\bar{A}}-Y_{A,\rm eq}Y_{\bar{A},\rm eq}\right) + {\langle \sigma v \rangle}_{SS \rightarrow A\bar{A}} \left(Y_S^2-\frac{Y_{\bar{A}}^2+CY_{\bar{A}}}{Y_{A,\rm eq}Y_{\bar{A},\rm eq}}Y_{S,\rm eq}^{2}\right)\, ,
\nonumber\\
 \frac{H^{\prime}x}{s}\frac{dY_S}{dx} &= 
-{\langle \sigma v \rangle}_{SS \rightarrow A\bar{A}} \left(Y_S^2-\frac{Y_{\bar{A}}^2+CY_{\bar{A}}}{Y_{A,\rm eq}Y_{\bar{A},\rm eq}}Y_{S,\rm eq}^{2}\right)
-{\langle \sigma v \rangle}_{SS \rightarrow A\bar{A}} \left(Y_S^2-\frac{Y_A^2-CY_A}{Y_{A,\rm eq}Y_{\bar{A},\rm eq}}Y_{S,\rm eq}^{2}\right). 
\end{align}

for $m_{A}<m_{S}$, where, $H^{\prime}=H\bigg[1+\bigg(\frac{x_R}{x}\bigg)^n\bigg]^{\frac{1}{2}}$, with $x=m_A/T,~x_R=m_A/T_R$ ($x=m_S/T,~x_R=m_S/T_R$) for $m_A>m_S$ ($m_A<m_S$). Therefore, solving for Eq.~(\ref{eq:boltzmann6}) or  Eq.~(\ref{eq:boltzmann7}), one can obtain DM relic abundances under the effect of non-standard evolution of the universe.

\section{Relic density calculations}
\label{res}
In this section, we solve for the multi-component Boltzmann equations for the asymmetric and symmetric dark matter described in Sec.~\ref{BE}. As discussed, we consider two cases leading to two different sets of Boltzmann equations with  A) $m_{A}>m_{S}$ and B) $m_{A}<m_{S}$ along with the assumption that the heavier DM candidate has no other annihilation except the hidden sector annihilation into lighter DM candidate. Considering the standard radiation dominating universe in effect, the choices lead to two different sets of parameters necessary for BE solutions given as

\bea
\label{par}
m_{A} , m_{S} ,  {\langle \sigma v \rangle}_{A\bar{A} \rightarrow SS},  {\langle \sigma v \rangle}_{S S \rightarrow XX}, C \hskip 5mm {\rm for} \hskip 5mm m_{A}>m_{s} \nonumber \\
m_{A} , m_{S} ,  {\langle \sigma v \rangle}_{SS \rightarrow A\bar{A}},  {\langle \sigma v \rangle}_{A\bar{A} \rightarrow XX}, C \hskip 5mm {\rm for} \hskip 5mm m_{A}<m_{s}\, . 
\eea
In the case of the non-standard evolution of the universe described in Sec.~\ref{NS}, two additional parameters get involved for the solution of BEs, which are $n$ and $T_R$. We first solve BEs in the case of the radiation domination era (Sec.~\ref{BE}) and further extend our analysis with the nonstandard cosmological era (Sec.~\ref{NS}) for comparison in detail.

\subsection{$m_{A}>m_{S}$}

\begin{figure}
    \begin{center}
        \includegraphics[width=0.45\textwidth]{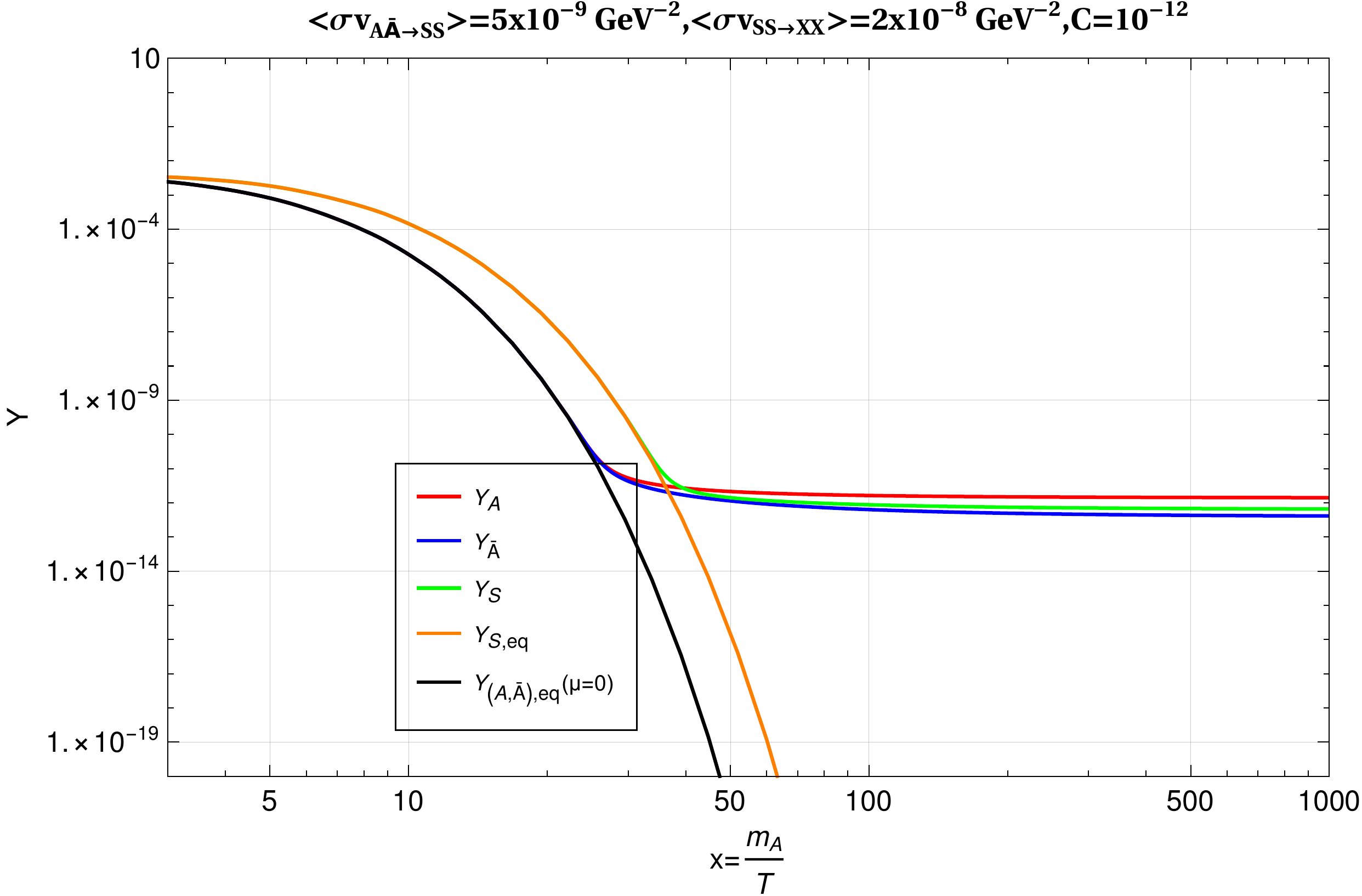}
        \includegraphics[width=0.45\textwidth]{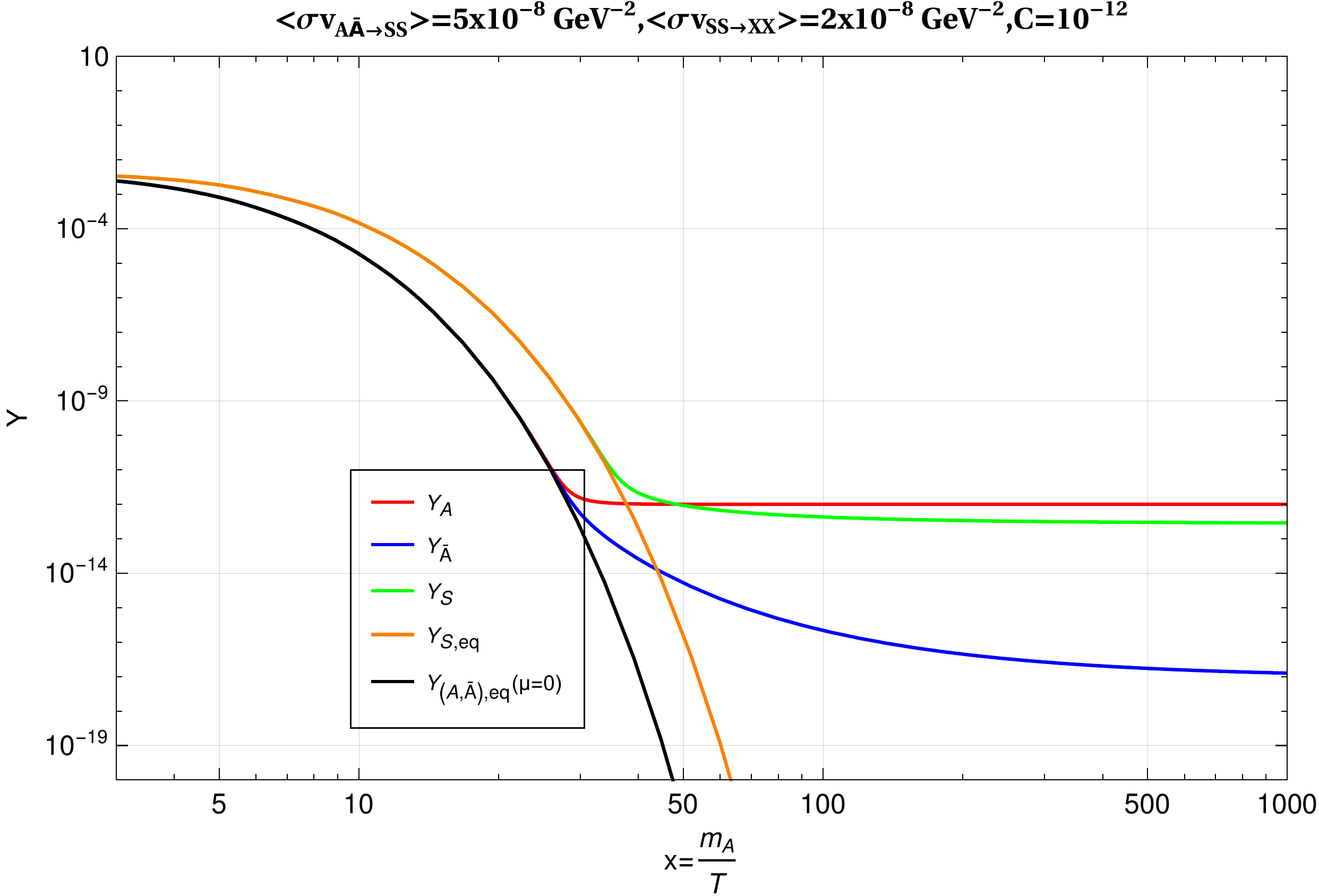}\\
        \includegraphics[width=0.45\textwidth]{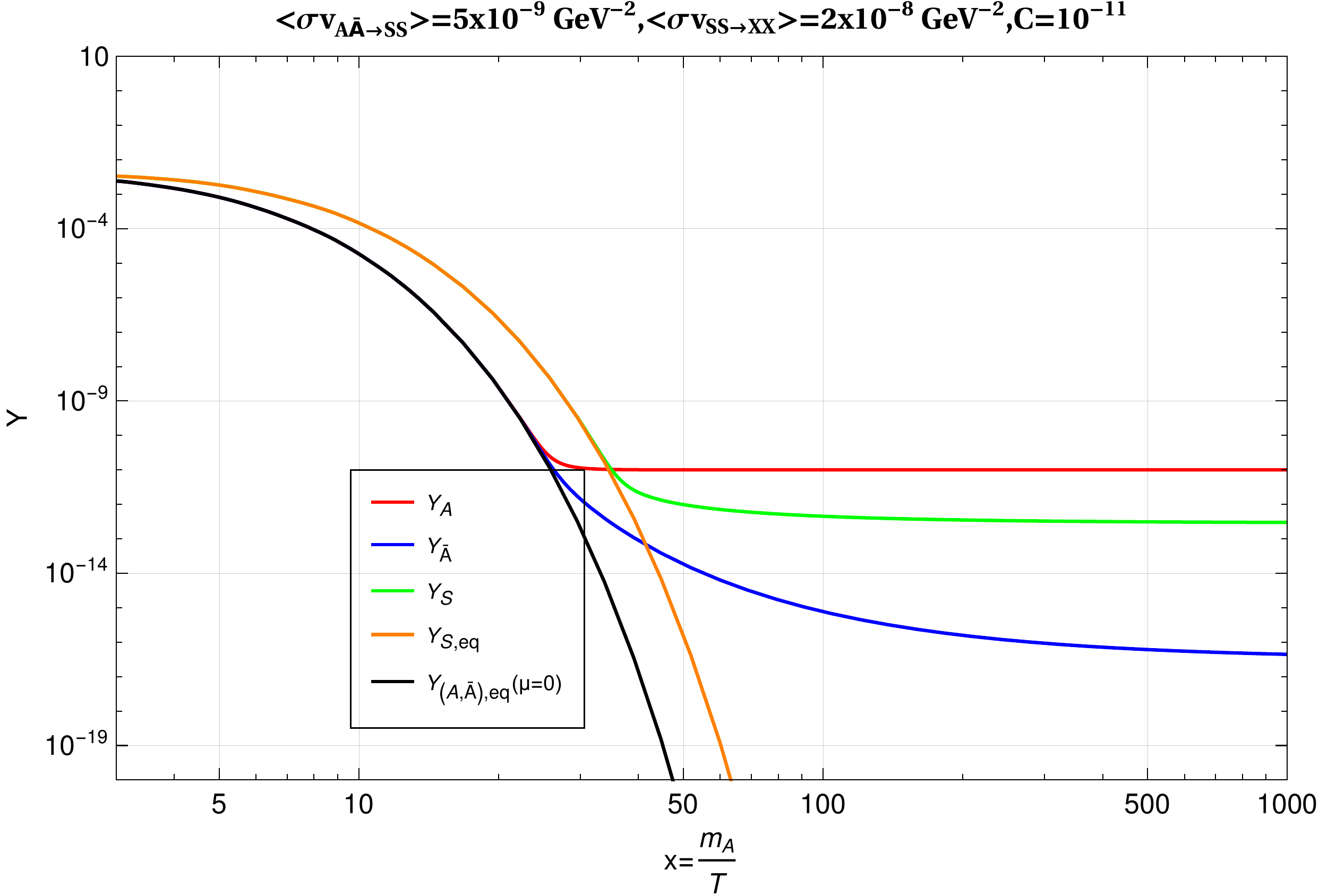}
        \includegraphics[width=0.45\textwidth]{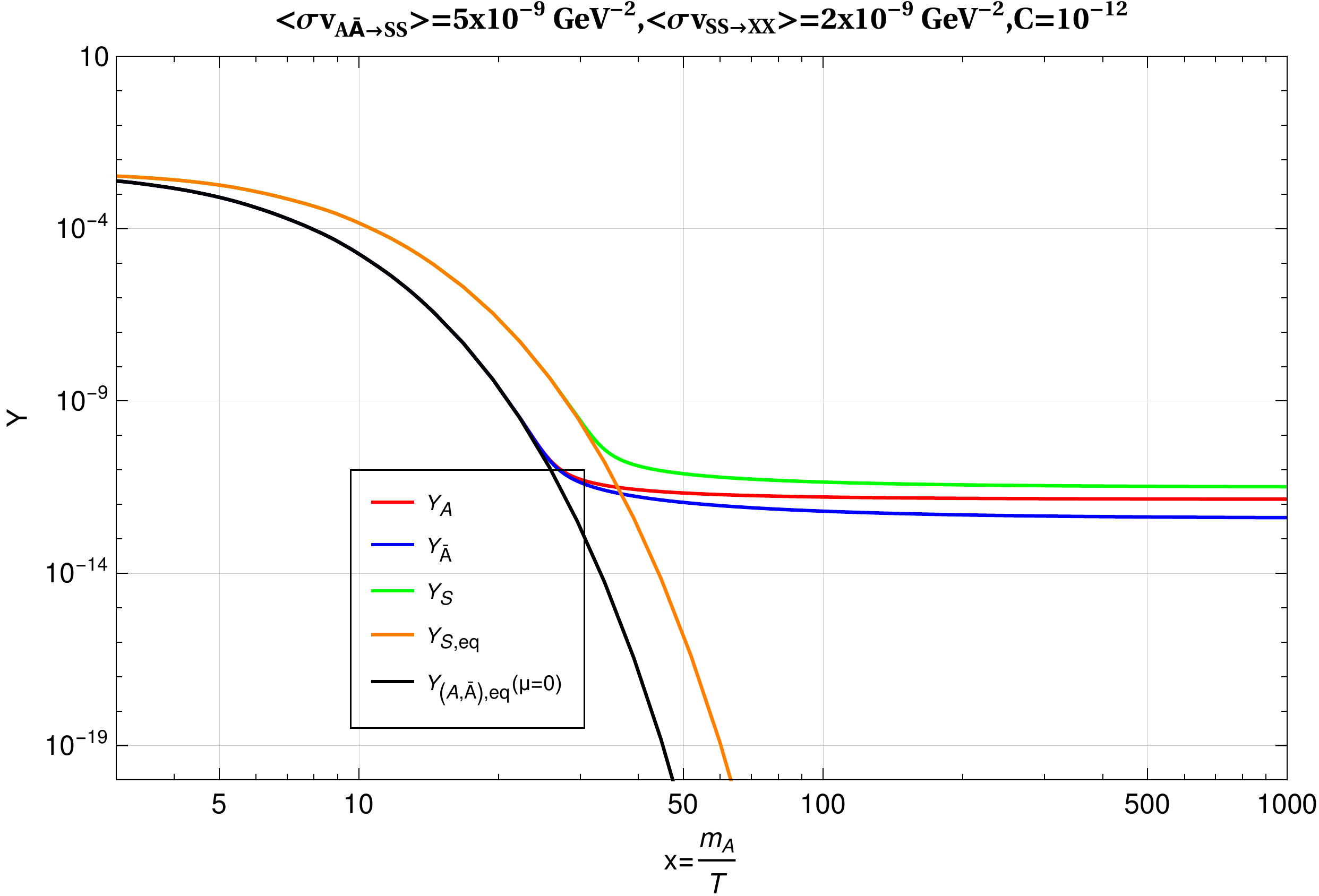}
            \caption{Evolution of co-moving number densities of DM candidates for benchmark data set in Table. I.}
        \label{fig1}
    \end{center}
\end{figure}

We begin with the calculation of dark matter abundance in a radiation-dominated era solving for coupled Boltzmann equation Eq.~(\ref{eq:boltzmann2}). Solutions to comoving DM abundances can be obtained for five parameters as mentioned in Eq.~(\ref{par}). For simplicity, initially, we consider the simplest case with the variation of only three parameters
${\langle \sigma v \rangle}_{A\bar{A} \rightarrow SS},  {\langle \sigma v \rangle}_{S S \rightarrow XX}, C$ for a fixed mass $m_A$ and $m_S$.
For this purpose, we assume for benchmark set values given in Table~\ref{t1}
and obtain the relic abundances for the given set of parameters and benchmark points (BPs). Respective variations of the comoving density of DM candidates are shown in Fig.~\ref{fig1}. From the Table~\ref{t1} (and Fig.~\ref{fig1}), we notice that increase in the hidden sector annihilation cross-section (dark annihilation) ${\langle \sigma v \rangle}_{A\bar{A} \rightarrow SS}$ reduce abundances of DM particles $A,~\bar{A}$ and $S$. In the case of  BP II, it is observed that the abundance of $\bar{A}$ gets reduced significantly, while its conjugate particle $A$ and the symmetric component $S$ suffer a mild reduction in abundance concerning BP I. An increase in the asymmetry parameter $C$ has almost a similar effect on the abundances of $\bar{A}$ and $S$, but conversely, it enhances the yield of $A$ when compared with BP I. This nature is evident, since as the abundance of $\bar{A}$ gets reduced, its contribution to the enhancement of the abundance of $S$ becomes negligible.
Finally, changing the annihilation ${\langle \sigma v \rangle}_{S S \rightarrow XX}$ (BP IV) only affects the abundance of $S$ particle as it is the annihilation into the visible sector, with no influence in the interactions of asymmetric dark matter component.

\begin{table}[htb]
 \begin{tabular}{| c | c | c | c | c | c | c | c |c|} 
 \hline
 BP & $m_{A,{\bar{A}}}$ GeV & $m_S$ GeV& ${\langle \sigma v \rangle}_{A\bar{A} \rightarrow SS}$ GeV$^{-2}$&  ${\langle \sigma v \rangle}_{SS \rightarrow XX}$ GeV$^{-2}$  &  $C$ &
 $\Omega_{A}h^2$ & $\Omega_{\bar{A}}h^2$ & $\Omega_{S}h^2$ \\ [0.5ex] 
 \hline
I  & 200 & 150 & 5$\times 10^{-9}$  & 2$\times 10^{-8}$ & $10^{-12}$ & 0.036 & 0.0110 & 0.013 \\

II & 200 & 150 & 5$\times 10^{-8}$ & 2$\times 10^{-8}$ & $10^{-12}$  & 0.025 & 4.3$\times10^{-7}$ & 0.005 \\

III  & 200 & 150 & 5$\times 10^{-9}$ & 2$\times 10^{-8}$ & $10^{-11}$ & 0.248 & 1.5$\times10^{-6}$ & 0.006 \\

IV  & 200 & 150 & 5$\times 10^{-9}$ & 2$\times 10^{-9}$ & $10^{-12}$ & 0.036 & 0.011 & 0.062 \\
 \hline
 \end{tabular}
\caption{Benchmark points with different parameters and respective DM abundances for $m_{A}>m_{S}$.}
\label{t1}
\end{table}

\begin{figure}
    \begin{center}
        \includegraphics[width=0.55\textwidth]{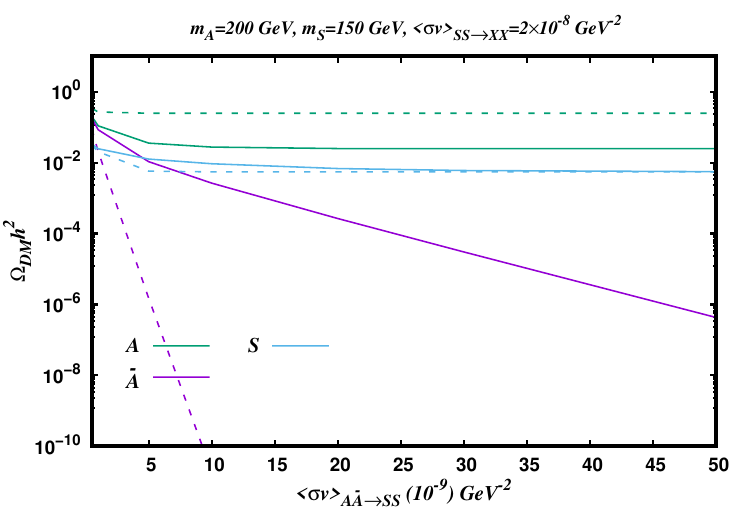}
            \caption{Variation of DM abundances with dark annihilation ${\langle \sigma v \rangle}_{A\bar{A} \rightarrow SS}$.}
        \label{fig2}
    \end{center}
\end{figure}

Although we obtain some insightful information about the dynamics of abundances of multi-particle asymmetric and symmetric DM systems, a broader overview is required to explore the effects of ``dark annihilation". For this purpose, we investigate the variations of
DM relic abundance with ${\langle \sigma v \rangle}_{A\bar{A} \rightarrow SS}$ for fixed $m_A,~m_S$ and ${\langle \sigma v \rangle}_{S S \rightarrow XX}$. Fig.~\ref{fig2} shows the variation of $\Omega_{DM}h^2$ plotted against dark annihilation cross-section ${\langle \sigma v \rangle}_{S S \rightarrow XX}$ for both asymmetric and symmetric DM particle taken into account. Solid (dashed) colored lines correspond to the asymmetry parameter value $C=10^{-12}$ ($C=10^{-11}$). We observe a decline in DM abundance with
increased dark sector annihilation. With increased ${\langle \sigma v \rangle}_{S S \rightarrow XX}$, the comoving abundance of $\bar{A}$ falls
significantly resulting in less production of symmetric dark matter candidate. This effect is reflected in the plots of Fig.~\ref{fig2}.
Enhancement in asymmetry parameter $C$ results in an increase in the hierarchy between the relic abundances of $A$ and $\bar{A}$ by lifting the abundance of $A$ and diminishing the abundance of $\bar{A}$.
The change in the parameter $C$ has no significant effect on the abundance of $S$. This is obvious from the Eq.~(\ref{eq:boltzmann2}) as the BE for $S$ depends on the term
$-{\langle \sigma v \rangle}_{A\bar{A} \rightarrow SS}C(Y_{A}-Y_{\bar{A}})=-{\langle \sigma v \rangle}_{A\bar{A} \rightarrow SS}C^2$. As a result, change in the relic abundance of $S$ is mostly governed by the annihilation of
$A~\bar{A} \rightarrow SS$ and the process $SS\rightarrow XX$.

\subsection{$m_{A}<m_{S}$}

\begin{figure}
    \begin{center}
        \includegraphics[width=0.45\textwidth]{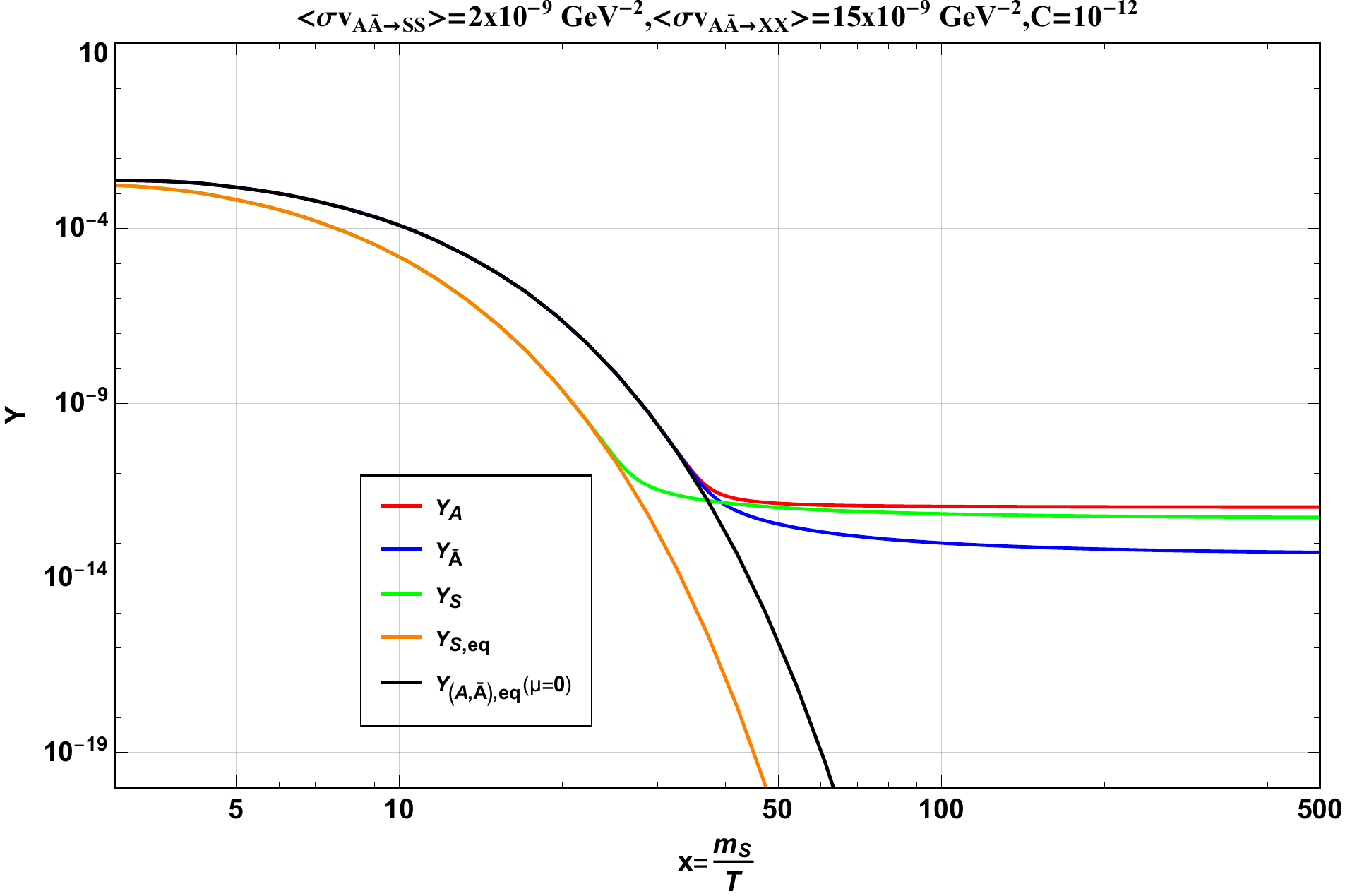}
        \includegraphics[width=0.45\textwidth]{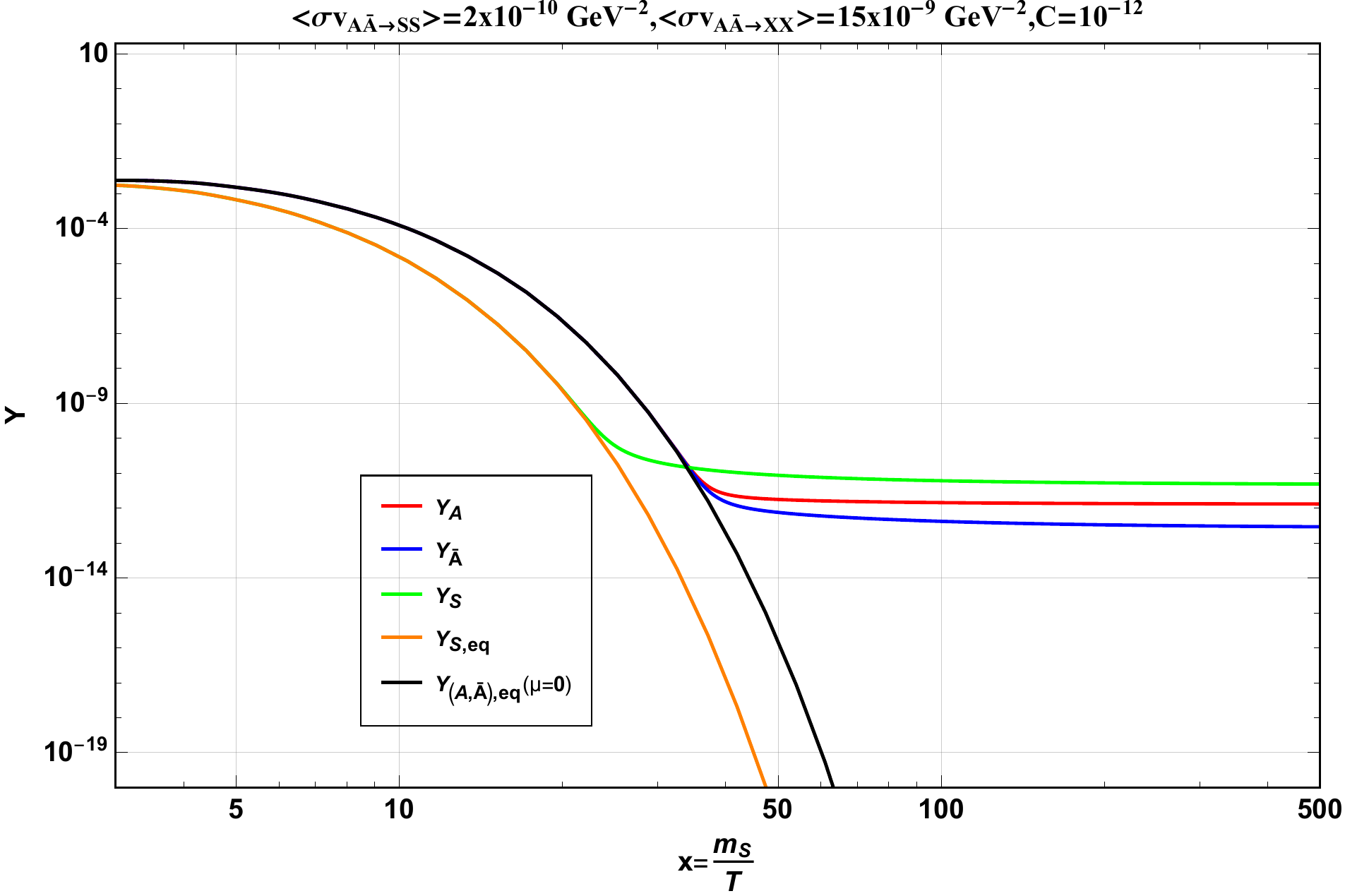}\\
        \includegraphics[width=0.45\textwidth]{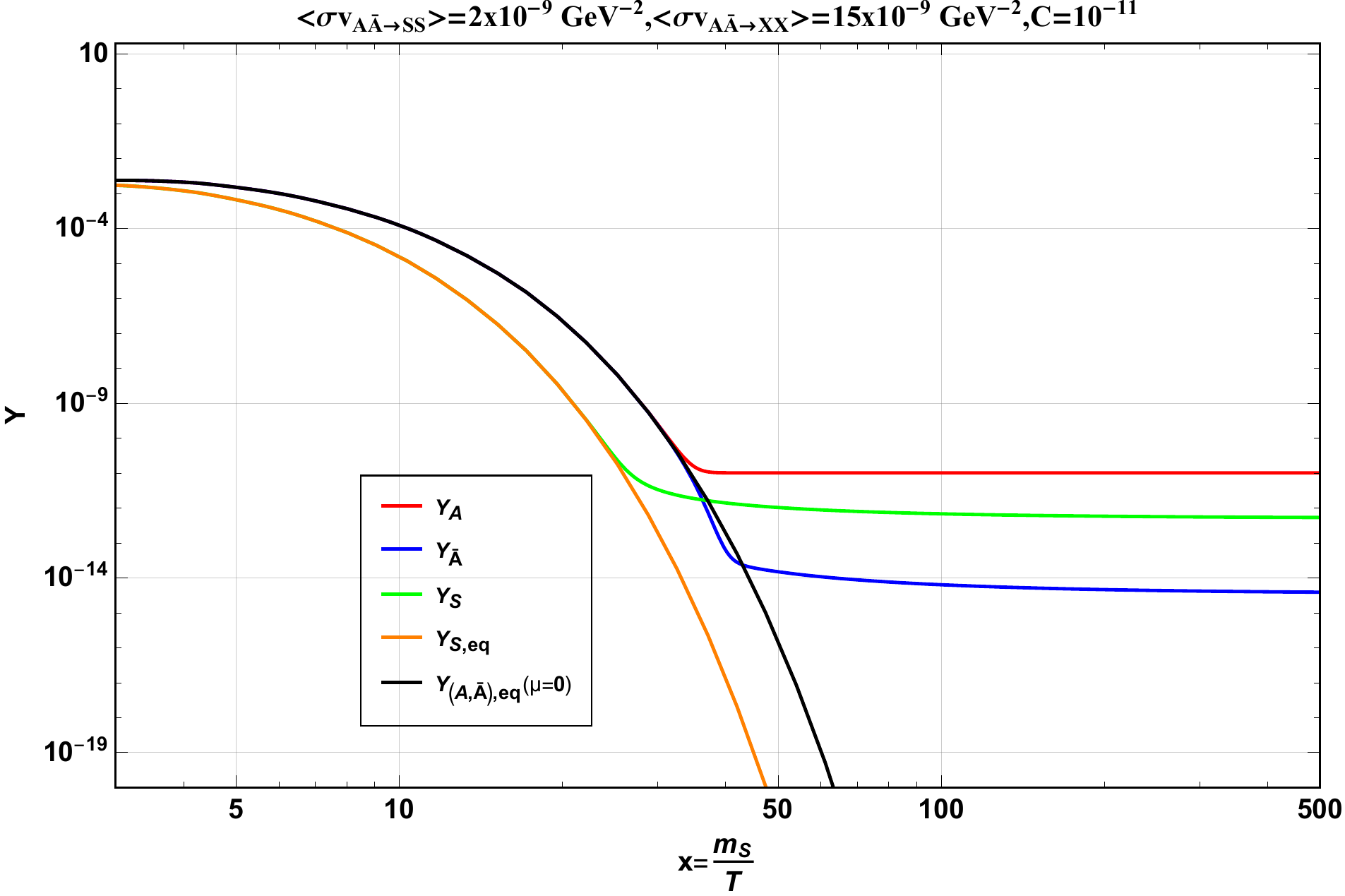}
        \includegraphics[width=0.45\textwidth]{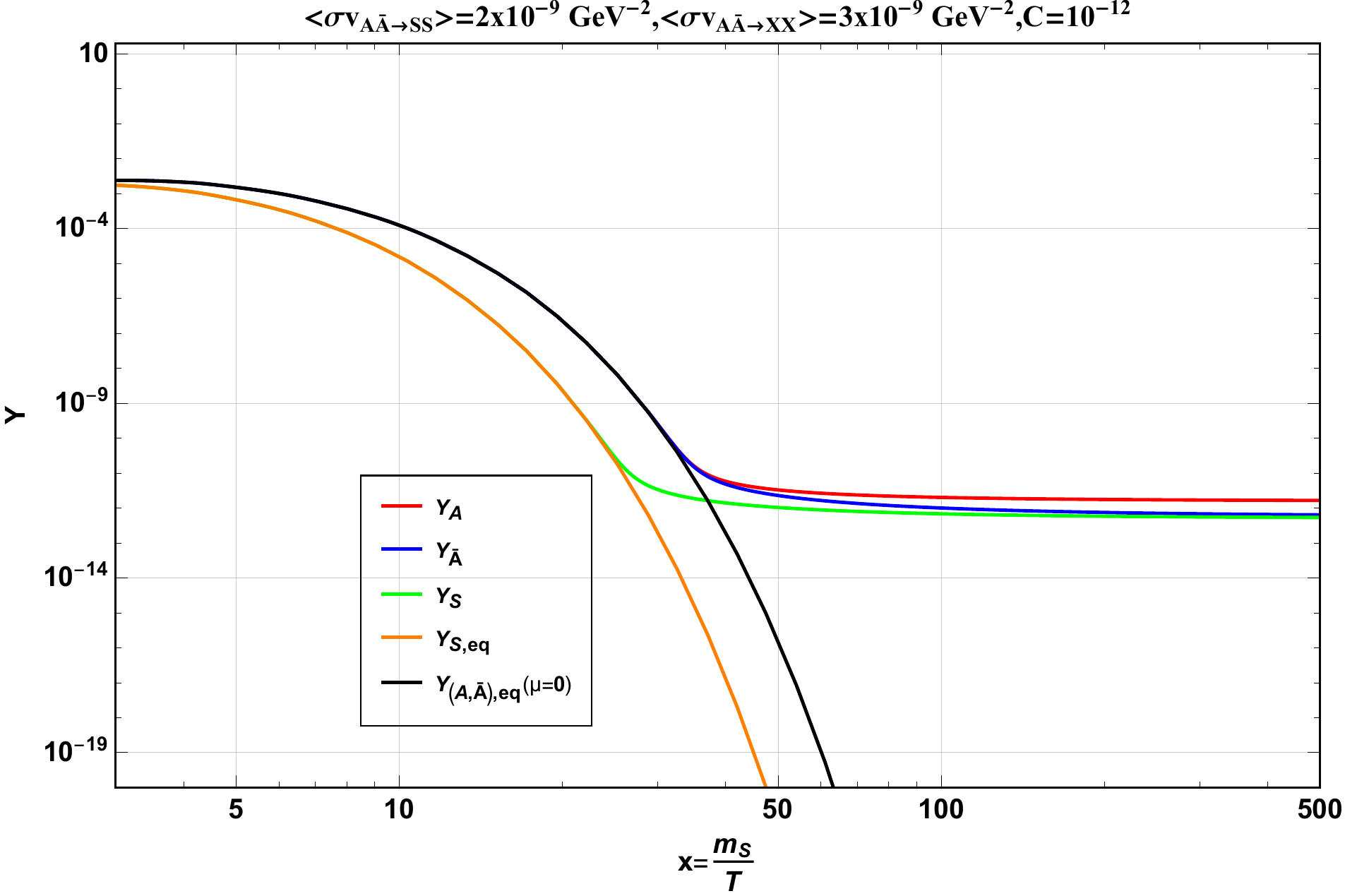}
            \caption{Same as Fig.~\ref{fig1} using benchmark data set in Table. II.}
        \label{fig3}
    \end{center}
\end{figure}

We now delve into the situation with $m_{A}<m_{S}$ in the standard radiation-dominated universe and solve for Eq.~(\ref{eq:boltzmann5}) with five sets of parameters mentioned in Eq.~(\ref{par}).
Similar to the case $m_{A}>m_{S}$, we consider a set of four benchmark points with fixed $m_{A}$ and $m_{S}$ and change the parameters ${\langle \sigma v \rangle}_{SS \rightarrow A\bar{A}},~{\langle \sigma v \rangle}_{A\bar{A} \rightarrow XX}$ and $C$. Evolution of comoving DM number density for $A,~\bar{A},~S$ using the four BPs mentioned in Table~\ref{t2} are demonstrated in Fig.~\ref{fig3}. Comparing the results BP I with BP II, we observe reduction (enhancement) in $Y_{i},~i=A,\bar{A},~S$ with increase (decrease) in the dark sector annihilation ${\langle \sigma v \rangle}_{SS \rightarrow A\bar{A}}$.
This is reflected in the values of DM abundances shown in Table~\ref{t2}.
An increase in the asymmetry parameter $C$ is found to increase the the $Y_{A}$
value while reducing the $Y_{\bar{A}}$ density. It is also observed that change in $C$
has no significant effect on the freeze out of symmetric DM candidate $S$ as we compare the results of BP I and BP III. This finding is also consistent as in Eq.~(\ref{eq:boltzmann5}), we encounter a term ${\langle \sigma v \rangle}_{SS \rightarrow A\bar{A}} \frac{C^2Y^2_{S,\rm{eq}}}{Y_{A,\rm eq},Y_{\bar{A},\rm eq}}$. We further compare the results of BP I and BP IV in Table~\ref{t2}, only to reveal that increase (decrease) in ${\langle \sigma v \rangle}_{A\bar{A} \rightarrow XX}$ only reduces (enhances) abundances of $A$ and $\bar{A}$, leaving the $S$ abundance unharmed.

\begin{table}[htb]
 \begin{tabular}{| c | c | c | c | c | c | c | c |c|} 
 \hline
 BP & $m_{A,{\bar{A}}}$ GeV & $m_S$ GeV& ${\langle \sigma v \rangle}_{SS \rightarrow A\bar{A}}$ GeV$^{-2}$&  ${\langle \sigma v \rangle}_{A\bar{A} \rightarrow XX}$ GeV$^{-2}$  &  $C$ &
 $\Omega_{A}h^2$ & $\Omega_{\bar{A}}h^2$ & $\Omega_{S}h^2$ \\ [0.5ex] 
 \hline
I  & 300 & 400 & 2$\times 10^{-9}$  & 15$\times 10^{-9}$ & $10^{-12}$ & 0.039 & 0.002 & 0.026 \\

II & 300 & 400 & 2$\times 10^{-10}$ & 15$\times 10^{-9}$ & $10^{-12}$  & 0.048 & 0.011 & 0.238 \\

III  & 300 & 400 & 2$\times 10^{-9}$ & 15$\times 10^{-9}$ & $10^{-11}$ & 0.373 & 0.001 & 0.026 \\

IV  & 300 & 400 & 2$\times 10^{-9}$ & 3$\times 10^{-9}$ & $10^{-12}$ & 0.061 & 0.023 & 0.026 \\
 \hline
 \end{tabular}
\caption{Benchmark points with different parameters and respective DM abundances for $m_{A}<m_{S}$.}
\label{t2}
\end{table}

\begin{figure}
    \begin{center}
        \includegraphics[width=0.55\textwidth]{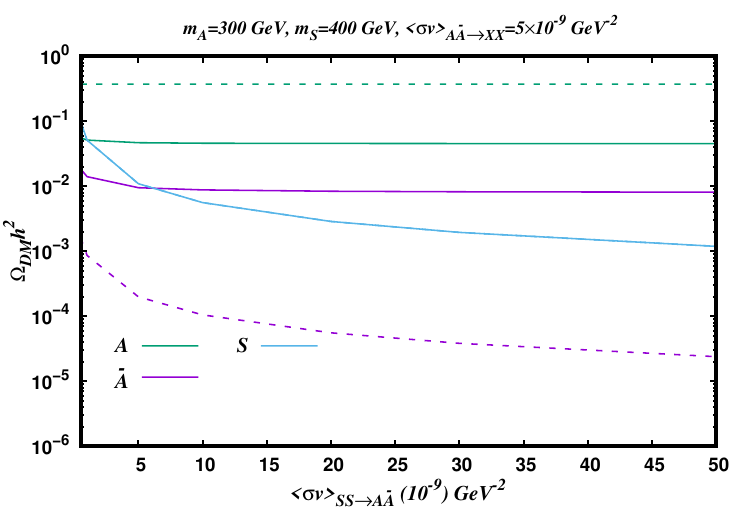}
                \caption{Variation of DM abundances with dark annihilation ${\langle \sigma v \rangle}_{SS \rightarrow A\bar{A}}$.}
        \label{fig4}
    \end{center}
\end{figure}

Similar to the $m_{A}>m_{S}$ scenario, we now investigate the effects of dark annihilation ${\langle \sigma v \rangle}_{SS \rightarrow A\bar{A}}$ shown in Fig.~\ref{fig4} on the abundances of DM particles. We observe a mild reduction in the abundances of both asymmetric and symmetric candidates with the increased value of $SS\rightarrow A \bar{A}$ annihilation. This effect is noted for both the choices of $C=10^{-12}$ (solid curves) and $C=10^{-11}$ (dashed curves) shown in the plots of Fig.~\ref{fig4}. 
Since an increased ${\langle \sigma v \rangle}_{SS \rightarrow A\bar{A}}$ tends to reduce the abundance of $S$, the resulting abundances of $A$ and $\bar{A}$ being coupled to $S$ is also found to be reduced. However, similar to the case of $m_{A}>m_{S}$ shown in Fig.~\ref{fig2}, the increase in asymmetry parameter overall enhances the abundance of $A$ and scales down the abundance of its complex conjugate $\bar{A}$ but has no effect on the abundance of $S$.

\subsection{Comparison with NS cosmological evolution}

\begin{table}[htb]
 \begin{tabular}{| c | c | c | c | c | c | c | c |c|} 
 \hline
  & $m_{A,{\bar{A}}}$ GeV & $m_S$ GeV& ${\langle \sigma v \rangle}_{A\bar{A} \rightarrow SS}$ GeV$^{-2}$&  ${\langle \sigma v \rangle}_{SS \rightarrow XX}$ GeV$^{-2}$  &  $C$ &
 $\Omega_{A}h^2$ & $\Omega_{\bar{A}}h^2$ & $\Omega_{S}h^2$ \\ [0.5ex] 
 \hline
RD  & 200 & 150 & 5$\times 10^{-9}$  & 2$\times 10^{-8}$ & $10^{-12}$ & 0.036 & 0.011 & 0.013 \\

$n=1,~T_R=1$ GeV & 200 & 150 & 5$\times 10^{-9}$ & 2$\times 10^{-8}$ & $10^{-12}$  & 0.056 & 0.031 & 0.026 \\

$n=2,~T_R=1$ GeV  & 200 & 150 & 5$\times 10^{-9}$ & 2$\times 10^{-8}$ & $10^{-12}$ & 0.076 & 0.051 & 0.040 \\
 \hline
 \end{tabular}
\caption{Same as BP-I of Table. I with non-standard cosmological parameters for $m_{A}>m_S$.}
\label{t3}
\end{table}

\begin{table}[htb]
 \begin{tabular}{| c | c | c | c | c | c | c | c |c|} 
 \hline
  & $m_{A,{\bar{A}}}$ GeV & $m_S$ GeV& ${\langle \sigma v \rangle}_{SS \rightarrow A\bar{A}}$ GeV$^{-2}$&  ${\langle \sigma v \rangle}_{A\bar{A} \rightarrow XX}$ GeV$^{-2}$  &  $C$ &
 $\Omega_{A}h^2$ & $\Omega_{\bar{A}}h^2$ & $\Omega_{S}h^2$ \\ [0.5ex] 
 \hline
RD  & 300 & 400 & 2$\times 10^{-9}$  & 15$\times 10^{-9}$ & $10^{-12}$ & 0.039 & 0.002 & 0.026 \\

$n=1,~T_R=1$ GeV & 300 & 400 & 2$\times 10^{-9}$ & 15$\times 10^{-9}$ & $10^{-12}$  & 0.051 & 0.013 & 0.069 \\

$n=2,~T_R=1$ GeV  & 300 & 400 & 2$\times 10^{-9}$ & 15$\times 10^{-9}$ & $10^{-12}$ & 0.075 & 0.038 & 0.138 \\
 \hline
 \end{tabular}
\caption{Same as BP-I of Table. II with non-standard cosmological parameters for $m_{A}<m_S$.}
\label{t4}
\end{table}

In this section, we consider the effects of the alternate cosmological history of the universe discussed in Sec.~\ref{NS} on the multi-particle dark matter evolution.
We use the modified BEs (Eqs.~(\ref{eq:boltzmann6}-\ref{eq:boltzmann7})) an obtain the abundances of asymmetric and symmetric DM candidates. For demonstration, we simply consider two benchmark points (BP I from Table~\ref{t1} and Table~\ref{t2}) and compare the relic abundances obtained for $n=1,2$ and $T_R=1$ GeV in Table~\ref{t3} and Table~\ref{t4} for $m_A>m_S$
and $m_A<m_S$. A smaller value of $T_R=1$ GeV ensures that the scalar $\phi$ field remains active during the evolution of DM candidates. Comparing with the results
from the standard radiation-dominated evolution of the universe, we find that the relic density of all the DM components gets enhanced with an increase in the value of $n$. This result is also consistent as a larger value of $n$ induces earlier freeze-out of DM species. 

\begin{figure}
    \begin{center}
        \includegraphics[width=0.45\textwidth]{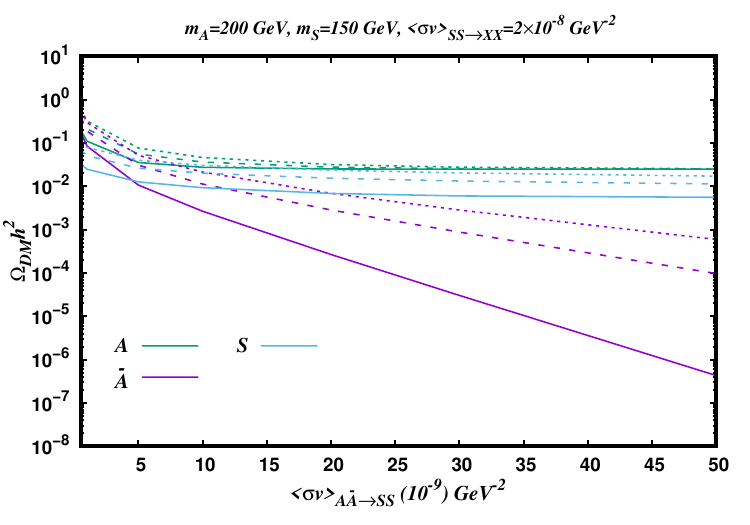}
        \includegraphics[width=0.45\textwidth]{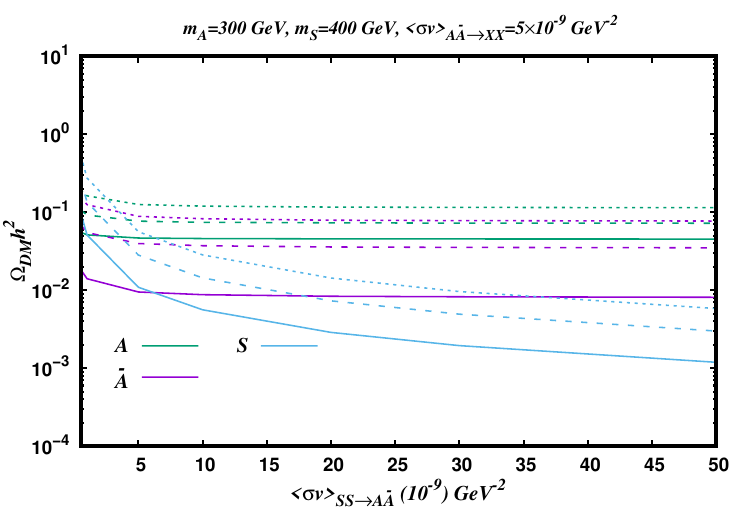}\\
            \caption{Same as Fig.3 and Fig. 4 with for variation of $n$ for $T_R=1$ GeV.}
        \label{fig5}
    \end{center}
\end{figure}

We further consider the effects of dark sector annihilation for the choices
$m_A>m_S$ and $m_A<m_S$ in the presence of non-standard expansion of the universe.
In Fig.~\ref{fig5}, we reproduce the results of Fig.~\ref{fig2} and Fig.~\ref{fig4} in the context of non-standard cosmology for $T_R=1$ GeV and  $n=1,2$ keeping other parameters fixed using $C=10^{-12}$ and compare the results with the standard radiation dominated era. Solid curves in the plots of Fig.~\ref{fig5} correspond to the usual radiation-dominated expansion of the universe.
Dashed (dotted) curves in Fig.~\ref{fig5} exhibit the variation of DM abundance with dark annihilation ${\langle \sigma v \rangle}_{A\bar{A}\rightarrow SS}$ ($m_A>m_S$) and ${\langle \sigma v \rangle}_{SS \rightarrow A\bar{A}}$ ($m_A<m_S$)  for $n=1$ ($n=2$) with $T_R=1$ GeV. For both the situations considered, we find an overall increase in relic abundances of all DM species.
This is obvious as with a faster expansion rate, DM species decouple earlier than expected than standard cosmology, and their abundance gets boosted due to
early freeze-out.

\begin{figure}
    \begin{center}
        \includegraphics[width=0.45\textwidth]{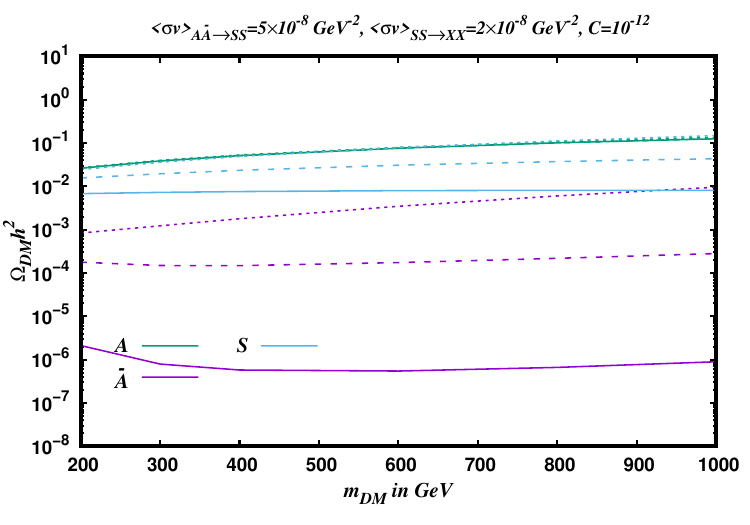}
        \includegraphics[width=0.45\textwidth]{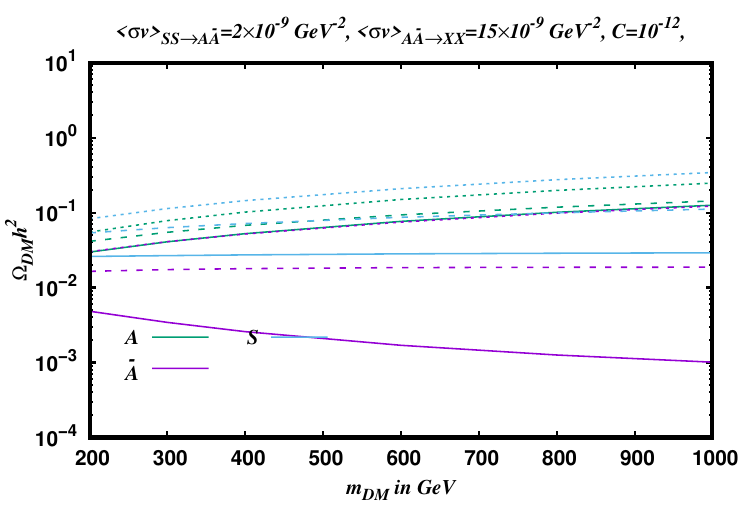}\\       
            \caption{Comparison of DM relic density against DM mass with the variation of $n$ for $T_R=1$ GeV. Left panel $m_A-m_S=10$ GeV, right panel $m_S-m_A=10$ GeV.}
        \label{fig6}
    \end{center}
\end{figure}

We now investigate the variation of DM relic abundance with masses of DM particles. We consider a mass splitting $m_A-m_S=10$ GeV between the $A$ and $S$ for $m_A>m_S$ and $m_S-m_A=10$ for $m_A<m_S$. In addition we consider a fixed set of annihilation cross-section ${\langle \sigma v \rangle}_{A\bar{A}\rightarrow SS}$, ${\langle \sigma v \rangle}_{S S \rightarrow XX}$ for $m_A>m_S$ and ${\langle \sigma v \rangle}_{SS \rightarrow A\bar{A}}$, ${\langle \sigma v \rangle}_{A\bar{A} \rightarrow XX}$ for $m_A<m_S$ and plot the variation of DM abundance with DM mass for fixed asymmetry parameter $C=10^{-12}$ in Fig.~\ref{fig6}. We use the same color and line type as in Fig.~\ref{fig5} to differentiate between the radiation-dominated era (solid curves) and the influence of alternate cosmology with $n=1$ (dashed curves) and $n=2$ (dotted curves). Relic abundances for non-standard expansion are obtained at $T_R=1$ GeV. From both sets of figures, enhancement in DM abundance is noticed in the presence of NS cosmology with DM mass varying from 200 GeV to 1000 GeV.   

\begin{figure}
    \begin{center}
        \includegraphics[width=0.45\textwidth]{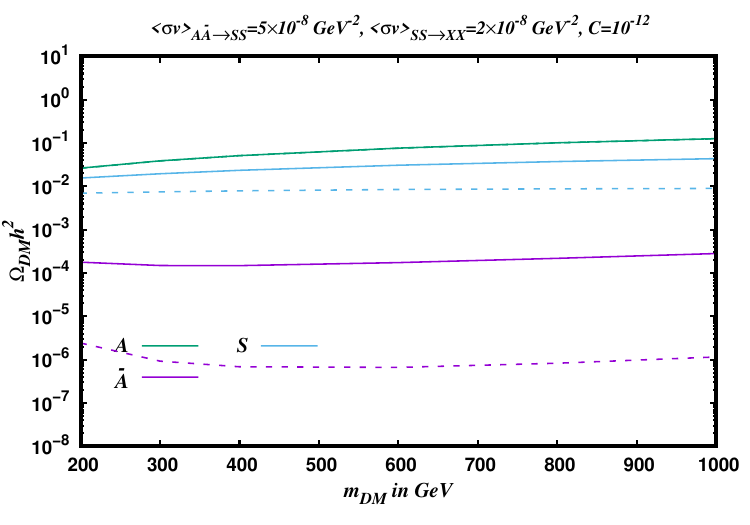}
        \includegraphics[width=0.45\textwidth]{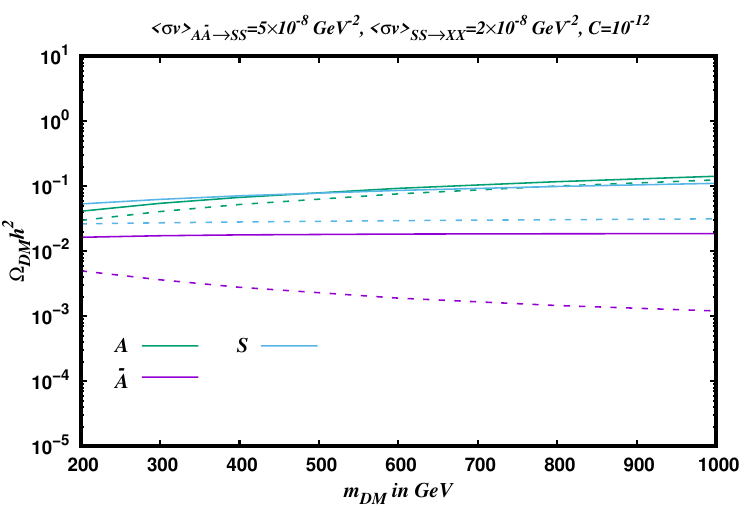}\\
        
            \caption{Same as Fig. 6, with varying $T_R$ and fixed $n~(=1)$ value.}
        \label{fig7}
    \end{center}
\end{figure}

Finally, we investigate the effect of the other non-standard cosmological parameter $T_R$. We reproduce the plots in Fig.~\ref{fig6} with the same set of parameters and $n=1$ and compare the changes in DM relic abundance for $T_R=1$ GeV (solid curves) and $T_R=100$ GeV (dashed curves). An increased value of $T_R$ indicates that the effect of the scalar field on the evolution of DM ceases at higher temperatures. As a result for a fixed value of $n$, relic abundances of both asymmetric and symmetric DM components get reduced. For larger $T_R$, DM relic densities will resemble the standard radiation-dominated evolution. This behaviour is realised in the plots of  Fig.~\ref{fig7}.

\section{Conclusions}
\label{con}

In this study, we explore a case of complex multi-particle dark matter dynamics 
with asymmetric and symmetric dark matter candidates. We derive the Boltzmann equations for the aforementioned two-component dark matter considering the heavier dark matter annihilates into lighter DM candidate and has no interaction with the visible sector. Therefore, one DM candidate is invisible to the SM sector having dark annihilation only. This brings up a new phenomenology as one of DM candidate freezes-out via dark annihilation. While the results are obtained for a standard radiation-dominated universe, we also compare our finding with the non-standard evolution of the universe leading to faster expansion. We summarise key characteristics of the work as follows

\begin{itemize}
\item For $m_A>m_S$, we observe that enhancement dark sector annihilation ($A \bar{A}\rightarrow SS$) results in a reduction in the abundance of all DM components. This occurs due to the combined reduction in abundances of asymmetric DM which couples with the annihilation cross-section in the Boltzmann equation. The visible sector annihilation $SS \rightarrow XX$ only affects the abundance of the $S$ component.

\item An increase in asymmetry parameter increases abundance of $A$ but $\bar{A}$ component suffers significant loss in abundance. Boltzmann equations reveal that the effect of the asymmetry parameter $C$ varies as $C^2$ for the symmetric DM $S$ and can be ignored safely. The effect in the abundance of $S$ occurs due to the changes in abundances of $A$ and $\bar{A}$, and due to change in annihilation cross-section $A\bar{A} \rightarrow SS$.

\item For $m_A<m_S$, dark annihilation $SS\rightarrow A \bar{A}$ and visible annihilation $A\bar{A} \rightarrow XX$ has similar effects on DM relic abundance as in the case of $m_A>m_S$.

\item With non-standard cosmology in effect, a large value of $n$, results in early freeze-out, and overall relic density of both symmetric and asymmetric DM tends to get boosted. 

\item In the presence of NS cosmology, large $T_R$ has a complementary effect on DM relic density, reducing the abundance for a chosen $n$.  

\end{itemize}
\vskip 5 mm 

In the present work, we have only explored the evolution of BEs with a model-independent approach. Direct and indirect detection are not discussed in the present work and will be considered in future work with a complete model. We have not discussed the origin of asymmetry $C$ for the asymmetric DM component. This purely depends on the chosen model and can originate via CP-violating interactions with some other beyond SM particle. It can further be generated from the dark annihilation itself when accompanied by loop-level interactions. Such studies may result in some interesting aspects of the model, such as the generation of matter-antimatter asymmetry, and also address other unresolved issues in the standard model of particle physics. We leave those interesting aspects of model-dependent studies for future works. There can be intriguing collider aspects of the present multi-particle asymmetric and symmetric DM that may provide some insight into hidden sector annihilation, which can be pursued in model-dependent studies. Apart from the above, the specific multi-particle DM scenario described in this work can also have various astrophysical and cosmological implications, that need to be addressed in future works. Finally, the influence of the non-standard evolution of the universe can further change the very outcomes for a specific model as discussed in the present article.

\vskip 3mm
\noindent {\bf Acknowledgments}: ADB acknowledges financial support from DST, India, under grant number IFA20-PH250 (INSPIRE Faculty Award).

\bibliographystyle{apsrev}
\bibliography{reference}

\end{document}